\documentclass[10pt,twocolumn,english,aps, prl,superscriptaddress,floatfix]{revtex4}
\usepackage[]{fontenc}
\usepackage[latin1]{inputenc}
\usepackage{graphicx}
\usepackage{amssymb}

\makeatletter

\usepackage{graphicx}
\usepackage{amssymb}

\usepackage{graphicx}
\usepackage{amssymb}

\usepackage{babel}

\usepackage{babel}
\makeatother
\begin{document}

\title{Effective charging energy for a regular granular metal array}
\author{Y.L.Loh}
\author{V.Tripathi}
\affiliation{Theory of Condensed Matter Group, Cavendish Laboratory, University
of Cambridge, Madingley Road, Cambridge CB3 0HE, United Kingdom}
\author{M.Turlakov}
\affiliation{R. Peierls Centre for Theoretical Physics, 1 Keble Road, Oxford
OX1 3NP, United Kingdom}

\begin{abstract}
We study the Ambegaokar-Eckern-Sch\"{o}n (AES) model for a regular array of metallic grains coupled by tunnel junctions of conductance $g$ and calculate both paramagnetic and diamagnetic terms in the Kubo formula for the conductivity.
We find analytically, and confirm by numerical path integral Monte Carlo methods, that for $0<g<4$ the conductivity obeys an Arrhenius law $\sigma(T)\sim\exp[-E^{*}(g)/T]$ with an effective charging energy $E^{*} (g)$ when the temperature is sufficiently low, due to a subtle cancellation between $T^2$ inelastic-cotunneling contributions in the paramagnetic and diamagnetic terms. 
We present numerical results for the effective charging energy and compare the
results with recent theoretical analyses. We discuss the different ways in which the experimentally observed $\sigma(T)\sim\exp[-\sqrt{T_{0}/T}]$ law could be attributed to disorder. 
\end{abstract}
\maketitle

The novel aspect of granular metals, which has attracted much experimental\cite{sheng73,chui81,simon87,gerber97,beverly02} and theoretical\cite{efetov02,efetov03,altland04} interest, is the interplay between  charging effects and incoherent intergrain hopping.  One of the main questions concerns the manner and extent to which intergrain tunneling delocalizes the charge in the low-temperature limit.
In the limit of $g\rightarrow 0,$ charging effects are dominant, implying an Arrhenius law for $\sigma(T)$.
Furthermore, even in the case of large $g$ and low temperatures, heuristic arguments for a one-dimensional chain based on the partition function of the AES model\cite{altland04} suggested an Arrhenius conductivity $\sigma(T)\sim\exp[-E^{*}(g)/T]$ at low temperatures with an effective charging energy $E^{*}(g)\sim E_{c}\exp[-\pi g/2].$
On the other hand, the conductance of a single quantum dot between two leads is dominated at low temperatures by inelastic cotunneling\cite{averin}, which gives a power-law temperature dependence.

We report several findings in this paper.  
First, for an \emph{ordered} granular array we find that even as $g$ becomes large, the conductivity in the AES model remains Arrhenius-like although the Coulomb gap is exponentially small in $g$, and therefore, the localization length of the charge shows no increase with decrease in temperature. This is despite the fact that intergrain correlation functions indeed have a power-law temperature dependence due to inelastic cotunneling processes. We explicitly trace this apparent paradox to a subtle cancellation of the anomalous power-law dependences in the Kubo formula for conductivity. This result is expected to hold for any dimensionality of the ordered array, and has been verified numerically for a 1D array for $0<g<4$. Our results are consistent with experiment\cite{beverly02} and with recent  theoretical work\cite{altland04,arovas}.  
Second, in view of the recent theoretical differences in the literature regarding the magnitude of $E^{*}(g)$, we also report the first numerical estimates for $E^{*}(g)$ for a regular array, extracted directly from the Kubo formula, which we expect will help resolve the theoretical issues. Finally, we also discuss the manner in which strong disorder in the grains' background potential and in intergrain tunneling conductance affect our conclusions,
especially in light of the experimentally observed soft-activation behavior
$\sigma(T)\sim\exp[-\sqrt{T_{0}/T}]$ in disordered granular systems. 

The Ambegaokar-Eckern-Sch\"{o}n (AES) model\cite{ambegaokar82,efetov02,efetov03}
is a valid description of a granular metal at temperatures higher than $\text{max}(g\delta, \delta)$ where $\delta$ is the mean level spacing at the Fermi energy in each grain.
We begin with the AES model for a regular array of metallic grains, \begin{eqnarray}
S[\varphi] & = & S_{c}[\varphi]+S_{t}[\varphi]=\frac{1}{4E_{c}}\sum_{\text{{\textbf{x}}}}\int_{\tau}(\partial_{\tau}\varphi_{\tau\text{{\textbf{x}}}})^{2}+\nonumber \\
 & + &
 \pi g
\sum_{\text{{\textbf{xa}}}}\int_{\tau\tau'}\alpha_{\tau-\tau'}\sin^2  \frac{\varphi_{\tau\text{{\textbf{x}}a}}-\varphi_{\tau'\text{{\textbf{xa}}}}}{2},\label{AES1}
\end{eqnarray}
 where $\tau$ is imaginary time, $\text{{\textbf{x}}}$ labels the grain positions, $\text{{\textbf{a}}}$
is a unit lattice vector, $\varphi_{\tau\text{{\textbf{xa}}}}=\varphi_{\tau\text{{\textbf{x}}}}-\varphi_{\tau\text{{\textbf{x+a}}}},$
and the dissipation kernel $\alpha_{\tau}=T^{2}/\sin^{2}[\pi T(\tau+i\epsilon)].$
The Matsubara fields $\varphi_{\tau\text{{\textbf{x}}}}$ satisfy
bosonic boundary conditions, $\varphi_{\beta\text{{\textbf{x}}}}=2\pi k_{\text{{\textbf{x}}}}+\varphi_{0\text{{\textbf{x}}}},$
where $k_{\text{{\textbf{x}}}}\in\mathbb{Z}$ is the winding number
at site $\mathbf{x}.$ The model describes transport in a granular
metal as a competition of charging ($S_{c}[\varphi]$) and dissipative
intergrain tunneling ($S_{t}[\varphi]$).

For convenience we now define the \emph{site correlator} ${\mathcal{C}}_{\tau_{1}\tau_{2}}={\mathcal{C}}_{\tau_{1}-\tau_{2}}=\langle\cos(\varphi_{\tau_{1}\text{{\textbf{x}}}}-\varphi_{\tau_{2}\text{{\textbf{x}}}})\rangle$
and the \emph{bond correlator} $\Pi_{\tau_{1}\tau_{2}}=\Pi_{\tau_{1}-\tau_{2}}=\langle\cos(\varphi_{\tau_{1}\text{{\textbf{xa}}}}-\varphi_{\tau_{2}\text{{\textbf{xa}}}})\rangle.$ 

The electromagnetic response function ${\mathcal{K}}=d\langle j\rangle/dA$
is given by the Kubo formula. It consists of diamagnetic (${\mathcal{K}}^{d}$)
and paramagnetic (${\mathcal{K}}^{p}$) contributions (${\mathcal{K}}={\mathcal{K}}^{d}+{\mathcal{K}}^{p}$).
In units of $a^{2-d}(e^{2}/h),$\begin{eqnarray}
{\mathcal{K}}_{\tau}^{d} & = & \pi g\left[{\mathcal{D}}_{\tau_{0},\tau_{0}+\tau}-\delta(\tau')\int_{\tau'}{\mathcal{D}}_{\tau_{0},\tau_{0}+\tau'}\right],\nonumber \\
\text{where} & \text{} & {\mathcal{D_{\tau}}}=\alpha_{\tau}\Pi_{\tau},\label{Kdia(def)}\\
{\mathcal{K}}_{\tau}^{p} & = & \sum_{\text{{\textbf{x}}}}\left\langle \hat{j}_{\tau_{0},\mathbf{x}_{0},\mathbf{a}_{0}}\hat{j}_{\tau_{0}+\tau,\mathbf{x}_{0}+\mathbf{x},\mathbf{a}_{0}}\right\rangle ,\nonumber \\
\text{where} &  & \hat{j}_{\tau_{1},\mathbf{x},\mathbf{a}}=\pi g\int_{\tau}\alpha_{\tau-\tau_{1}}\sin(\varphi_{\tau\mathbf{xa}}-\varphi_{\tau_{1}\mathbf{xa}}),\label{Kpara(def)}\end{eqnarray}
 where $\Pi$ is the bond correlator of phases on adjacent grains
at different times, and $\tau_{0},\mathbf{x}_{0},\mathbf{a}_{0}$
may be chosen arbitrarily. The real part of the conductivity is related
to the imaginary part of the response function, \begin{eqnarray}
\text{Re }\sigma_{\omega} & = & \text{Im }\left[\frac{{\mathcal{K}}_{\omega}^{d}+{\mathcal{K}}_{\omega}^{p}}{\omega}\right].\label{conductivity}\end{eqnarray}

The Kubo formula Eq.(\ref{conductivity}) only gives the part of the
conductivity that is Ohmic and {}``extensive'', i.e., $\sigma=\lim_{L,A\rightarrow0}(GL/A),$
where $G=\lim_{V\rightarrow0}(I/V)$ is the zero bias conductance
of the sample of length $L$ and cross section area $A.$ In a specimen
of finite size, there are other contributions to the current (e.g.,
inelastic cotunneling) that are missed out by the Kubo formula.

Next we proceed to calculate the conductivity beginning with small
values of intergrain conductance $g.$ In this case, the conductivity
may be expanded in a perturbation series in $g.$ 
To the lowest order in $g$, only the diamagnetic part of the electromagnetic
response matters, and it is straightforward to show using Eq.(\ref{Kdia(def)})
and Eq.(\ref{conductivity}) that the d.c. conductivity is 
\begin{equation}
\frac{\sigma^{d}}{\sigma_{0}}\approx2e^{-\beta E_{c}}+2\beta E_{c}\,
e^{-2\beta E_{c}}.\label{expconduct}\end{equation}
We now perform perturbation theory in $g$ for the site and bond correlators,
taking $S_{c}[\varphi]$ as the bare action and $S_{t}[\varphi]$
as the perturbation. The first-order correction to the site correlator
can be expressed in terms of bare correlators (the angle
brackets represent expectations under the bare action):
\begin{eqnarray}
\!\!\!\!\mathcal{C}_{\tau_{1}\tau_{2}\mathbf{x}_{1}}^{(1)}  & = & 
\frac{\pi g}{2}\sum_{\mathbf{x}_{3}\mathbf{a}_{3}}
\int\!\!\!\!\int_{\tau_{3}\tau_{4}}\alpha_{\tau_{3}-\tau_{4}}\times  \nonumber \\
 & \times & \left[\left\langle e^{i(\varphi_{\tau_{1}\mathbf{x}_{1}}-\varphi_{\tau_{2}\mathbf{x}_{1}})}e^{i(\varphi_{\tau_{3}\mathbf{x}_{3}\mathbf{a}_{3}}-\varphi_{\tau_{4}\mathbf{x}_{3}\mathbf{a}_{3}})}\right\rangle -\right.
\nonumber \\ & - & 
 \left.\left\langle e^{i(\varphi_{\tau_{1}\mathbf{x}_{1}}-\varphi_{\tau_{2}\mathbf{x}_{1}})}\right\rangle \left\langle e^{i(\varphi_{\tau_{3}\mathbf{x}_{3}\mathbf{a}_{3}}-\varphi_{\tau_{4}\mathbf{x}_{3}\mathbf{a}_{3}})}\right\rangle \right].\label{ctau(O1)}\end{eqnarray}

The first exponential in Eq.(\ref{ctau(O1)}) involves phases on grain
\(\mathbf{x}_{1}\) while the second exponential involves phases on
grains \(\mathbf{x}_{3}\) and \(\mathbf{x}_{4}=\mathbf{x}_{3}+\mathbf{a}_{3}.\)
The averages in Eq.(\ref{ctau(O1)}) can be factorized into averages over phases
on separate grains.

(a) If \(\mathbf{x}_{1},\,\mathbf{x}_{3}\,\,\text{and }\mathbf{x}_{4}\)
are all distinct, the two expectations cancel each other exactly. 

(b) If \(\mathbf{x}_{1}=\mathbf{x}_{3}\) (or equivalently, if \(\mathbf{x}_{1}=\mathbf{x}_{4}\)),
the integrand becomes 
$ (\mathcal{C}_{\tau_{1}\tau_{2}\tau_{3}\tau_{4}}-\mathcal{C}_{\tau_{1}\tau_{2}}
\mathcal{C}_{\tau_{3}\tau_{4}})\mathcal{C}_{\tau_{3}\tau_{4}},$  
in terms of the bare single-grain correlators \(\mathcal{C} \) 
and $\mathcal{C}_{\tau_1\tau_2\tau_3\tau_4}=
\left\langle
 e^{i(\varphi_{\tau_1}  - \varphi_{\tau_2} + \varphi_{\tau_3} - \varphi_{\tau_4}  )}
\right\rangle$.

$\mathcal{C}_{\tau_{1}\tau_{2}\tau_{3}\tau_{4}}$ 
takes its largest value, 1, when $\tau_1\approx\tau_2$ and $\tau_3\approx \tau_4$, or when $\tau_1\approx\tau_4$ and $\tau_3\approx \tau_2$; it decays exponentially with $|\tau_1-\tau_2|$, etc.  
This behavior is approximately described by Wick's theorem, $\mathcal{C}_{\tau_{1}\tau_{2}\tau_{3}\tau_{4}}\approx\mathcal{C}_{\tau_{1}\tau_{2}}\mathcal{C}_{\tau_{3}\tau_{4}}+\mathcal{C}_{\tau_{1}\tau_{4}}\mathcal{C}_{\tau_{3}\tau_{2}};$ however, when all the time indices are equal,
$\mathcal{C}_{\tau_{1}\tau_{2}\tau_{3}\tau_{4}}=1$ (not $2$).
We handle this by defining $\alpha_{\tau_{3}-\tau_{4}}^{\text{reg}}$ as $\alpha_{\tau_{3}-\tau_{4}}$ times a regularizing factor that vanishes at $\tau_{3}=\tau_{4}$:
\begin{eqnarray}
\mathcal{C}_{\tau_{1}\tau_{2}\mathbf{x}_{1}}^{(1)} & = & \frac{\pi gz}{2}\int\!\!\!\!\int_{\tau_{3}\tau_{4}}\alpha_{\tau_{3}-\tau_{4}}^{\text{reg}}\mathcal{C}_{\tau_{1}\tau_{4}}\mathcal{C}_{\tau_{3}\tau_{2}}\mathcal{C}_{\tau_{3}\tau_{4}}\nonumber \\
 & \approx & \frac{\pi gz}{2}\left(\frac{2}{E_{c}}\right)^{2}\alpha_{\tau_{1}-\tau_{2}}^{\text{reg}}\mathcal{C}_{\tau_{1}\tau_{2}}.\label{ctau(O1)2}\end{eqnarray}
 where $z$ is the coordination number of the grain. This is exponentially
small ($\sim\exp[-E_{c}|\tau_{1}-\tau_{2}|]$). Thus to first order
in $g,$ the exponential decay of the site correlator is not changed,
in contrast with the case of the bond correlator that we now discuss.

The first correction to the bond correlator is
\begin{eqnarray}
\!\!\!\!\!\!\!\!\! \mathcal{C}_{\tau_{1}\tau_{2}\mathbf{x}_{1}\mathbf{a}}^{(1)}  & = & 
\frac{\pi
  g}{2}\sum_{\mathbf{x}_{3}\mathbf{a}_{3}}\int\!\!\!\!\int_{\tau_{3}\tau_{4}}\alpha_{\tau_{3}-\tau_{4}} \times \nonumber \\
 & \!\!\!\!\!\!\!\!\!\!\times & \!\!\!\!\!\!\!\left[\left\langle e^{i(\varphi_{\tau_{1}\mathbf{x}_{1}\mathbf{a}_{1}}-
\varphi_{\tau_{2}\mathbf{x}_{1}\mathbf{a}_{1}})}
e^{i(\varphi_{\tau_{3}\mathbf{x}_{3}\mathbf{a}_{3}}-
\varphi_{\tau_{4}\mathbf{x}_{3}\mathbf{a}_{3}})}\right\rangle-\!\!\!\!\right.\nonumber
\\ 
& \!\!\!\!\!\!\!\!\!\!- & \!\!\!\!\!\!\!\!\!
\left.\left\langle
e^{i(\varphi_{\tau_{1}\mathbf{x}_{1}\mathbf{a}_{1}}-\varphi_{\tau_{2}\mathbf{x}_{1}\mathbf{a}_{1}})}\right\rangle\!\!\left\langle
e^{i(\varphi_{\tau_{3}\mathbf{x}_{3}\mathbf{a}_{3}}-\varphi_{\tau_{4}\mathbf{x}_{3}\mathbf{a}_{3}})}\right\rangle\right].
\label{cbond(O1)}\end{eqnarray}

\begin{figure}
\includegraphics[scale=0.9]{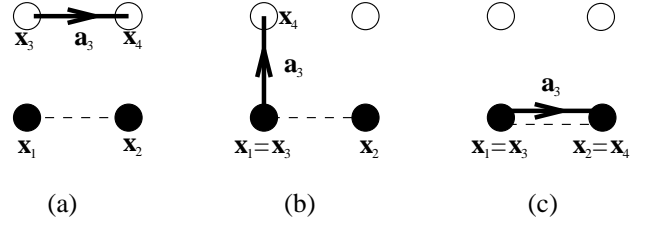}

\caption{\label{f:bondcorrelatordiagram} Contributions to the bond correlator
$\Pi_{\tau\mathbf{a}}.$ (a) No sites in common, (b) one common site,
and (c) two common sites.}
\end{figure}

The first exponential in Eq.(\ref{cbond(O1)}) involves phases on
grains $\mathbf{x}_{1}$ and $\mathbf{x}_{2}=\mathbf{x}_{1}+\mathbf{a}_{1},$
while the second exponential involves phases on grains $\mathbf{x}_{3}$
and $\mathbf{x}_{4}=\mathbf{x}_{3}+\mathbf{a}_{3}.$ This is illustrated
in Fig.\ref{f:bondcorrelatordiagram}.

(a) If the two bonds do not share a common site, the expectation of
the product of the cosines of the phases on these bonds factorizes.
Hence the two terms cancel and there is no expectation.

(b) If the two bonds have one site in common, such that $\mathbf{x}_{1}=\mathbf{x}_{3}$
for example, then the integrand becomes $({\cal C}_{\tau_{1}\tau_{2}\tau_{3}\tau_{4}}-{\cal C}_{\tau_{1}\tau_{2}}{\cal C}_{\tau_{3}\tau_{4}}){\cal C}_{\tau_{3}\tau_{4}}.$

(c) If the two bonds have both sides in common, i.e., $\mathbf{x}_{1}=\mathbf{x}_{3}$
and $\mathbf{x}_{2}=\mathbf{x}_{4}$ (and $\mathbf{a}_{1}=\mathbf{a}_{3}$),
then the integrand is $({\cal C}_{\tau_{1}\tau_{2}\tau_{3}\tau_{4}})^{2}-({\cal C}_{\tau_{1}\tau_{2}})^{2}({\cal C}_{\tau_{3}\tau_{4}})^{2}.$

The second term in case (c) is always small ($\sim\exp[-2E_{c}|\tau_{1}-\tau_{2}|]$),
but the first contains peaks when its four arguments can be partitioned
into even-odd pairs of $\tau$'s that are close to each other. Hence
when performing the integral over $\tau_{3}$ and $\tau_{4},$ one
finds an important contribution from the region where $|\tau_{1}-\tau_{4}|$
and $|\tau_{2}-\tau_{3}|$ are small:\begin{equation}
\mathcal{C}_{\tau_{1}\tau_{2}\mathbf{x}_{1}\mathbf{a}_{1}}^{(1c)}=\frac{\pi g}{2}\int\!\!\!\!\int_{\tau_{3}\tau_{4}}\!\!\!\!\alpha_{\tau_{3}-\tau_{4}}(\mathcal{C}_{\tau_{1}\tau_{2}\tau_{3}\tau_{4}})^{2}\approx\frac{2\pi g}{E_{c}^{2}}\alpha_{\tau_{1}-\tau_{2}}.\label{cbond(O1)2}\end{equation}
 In accordance with Griffiths' theorem \cite{griffiths,ginibre},
the long-ranged interaction produces a qualitative change in the behavior
of the correlation function.

Putting $\Pi_{\tau\mathbf{a}}$ into the expression for the diamagnetic
response $\mathcal{K}^{d}$ gives the second order correction to $\mathcal{K}^{d}:$\begin{eqnarray}
\mathcal{K}_{\tau}^{d(2)} & \approx & \frac{2\pi^{2}g^{2}}{E_{c}^{2}}\left[\alpha_{\tau}^{2}-\delta(\tau)\int_{\tau'}\alpha_{\tau'}^{2}\right].\label{kdia(O2)}\end{eqnarray}
 This contributes to the conductivity a term proportional to $g^{2}T^{2}$
at low temperatures --- a much weaker insulating behavior than both
Arrhenius ($\exp[-E_{c}/T]$) and soft activation ($\exp[-\sqrt{T_{0}/T}]$)
laws commonly encountered in experiment. This is reminiscent of inelastic
cotunneling processes, which are known to give a \emph{conductance}
proportional to $T^{2}$ for a single quantum dot\cite{averin}; however,
the conductivity of a granular metal is related to the conductance
of a \emph{macroscopic} specimen, and in order to contribute to this
conductance, $L$ electrons would have to cotunnel simultaneously
along each segment of the path linking one electron to the other,
which would give a negligible conductance proportional to $T^{2L}.$
In order to obtain the correct behavior one must also consider the
paramagnetic contribution.

The first term of $\mathcal{K}^{p}$ is also proportional to $g^{2}$
and can be calculated from the bare action. From Eq.(\ref{Kpara(def)}),\begin{eqnarray}
\mathcal{K}_{\tau_{1}\tau_{3}\mathbf{x}_{1}\mathbf{x}_{3}\mathbf{a}_{1}\mathbf{a}_{3}}^{p}=\pi^{2}g^{2}\int\!\!\!\!\int_{\tau_{2}\tau_{4}}\alpha_{\tau_{1}-\tau_{2}}\alpha_{\tau_{3}-\tau_{4}}\times\qquad\nonumber \\
\times\langle\sin(\varphi_{\tau_{1}\mathbf{x}_{1}\mathbf{a}_{1}}-\varphi_{\tau_{2}\mathbf{x}_{1}\mathbf{a}_{1}})\sin(\varphi_{\tau_{3}\mathbf{x}_{3}\mathbf{a}_{3}}-\varphi_{\tau_{4}\mathbf{x}_{3}\mathbf{a}_{3}})\rangle.\label{kpara(O1)1}\end{eqnarray}
 The first sine involves phases on grains $\mathbf{x}_{1}$ and $\mathbf{x}_{2}=\mathbf{x}_{1}+\mathbf{a}_{1},$
while the second sine involves the phases on grains $\mathbf{x}_{3}$
and $\mathbf{x}_{4}=\mathbf{x}_{3}+\mathbf{a}_{3}.$ Marking these
bonds on the lattice gives the same picture as Fig.\ref{f:bondcorrelatordiagram}.

(a) If these bonds do not share a common site, the average is zero
by symmetry.

(b) If the bonds share one common site, the average is still zero.

(c) The only finite contribution arises when these bonds have both
sites in common. In this case, the term in the angle brackets in Eq.(\ref{kpara(O1)1})
gives \begin{eqnarray}
\mathcal{K}_{\tau_{1}\tau_{3}\mathbf{x}_{1}\mathbf{x}_{3}\mathbf{a}_{1}\mathbf{a}_{3}}^{p} & = & -\frac{\pi^{2}g^{2}}{2}\int\!\!\!\!\int_{\tau_{2}\tau_{4}}\alpha_{\tau_{1}-\tau_{2}}\alpha_{\tau_{3}-\tau_{4}}\times\nonumber \\
 &  & \times[(\mathcal{C}_{\tau_{1}\tau_{2}\tau_{3}\tau_{4}})^{2}-(\mathcal{C}_{\tau_{1}\tau_{2}\tau_{4}\tau_{3}})^{2}].\label{kpara(O1)2}\end{eqnarray}
 When $|\tau_{1}-\tau_{2}|$ and $|\tau_{3}-\tau_{4}|$ are small,
the two correlators in Eq.(\ref{kpara(O1)2}) cancel out. When $|\tau_{1}-\tau_{4}|$
and $|\tau_{2}-\tau_{3}|$ are small, integration over $\tau_{2}$
and $\tau_{4}$ gives $-(2\pi^{2}g^{2}/E_{c}^{2})\alpha_{\tau_{1}-\tau_{3}}.$
When $|\tau_{1}-\tau_{3}|$ and $|\tau_{2}-\tau_{4}|$ are small,
performing the integration gives $(2\pi^{2}g^{2}/E_{c}^{2})\delta(\tau_{1}-\tau_{3})\int_{\tau'}\alpha_{\tau'}^{2}.$
Gathering these results, the correction to $\mathcal{K}_{\tau}^{p}$
at large $\tau$ is \begin{eqnarray}
\mathcal{K}_{\tau}^{p(2)} & \approx & -\frac{2\pi^{2}g^{2}}{E_{c}^{2}}\left[\alpha_{\tau}^{2}-\delta(\tau)\int_{\tau'}\alpha_{\tau'}^{2}\right].\label{kpara(O2)}\end{eqnarray}
 This is equal and opposite to the correction to $\mathcal{K}^{d},$
Eq.(\ref{kdia(O2)}). Hence the anomalous $T^{2}$ terms in the conductivity
cancel out when all contributions are taken into account. 

\begin{figure}
\includegraphics[%
  width=0.50\textwidth,
  keepaspectratio]{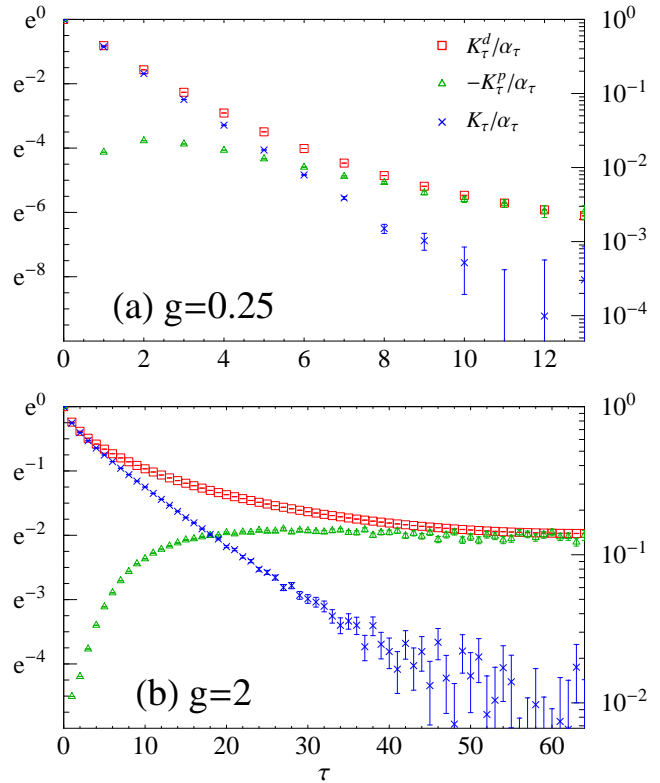}

\caption{\label{f:numerics} (Color online) Monte Carlo estimates of electromagnetic response functions (scaled by $\alpha_\tau$) for a ring of 16 grains with $E_c=\frac{1}{2}$ and $\beta=128$. 
The anomalous power-law contributions in ${\cal K}_{\tau}^{d}/\alpha_{\tau}$ and
${\cal K}_{\tau}^{p}/\alpha_{\tau}$ exactly cancel out in ${\cal K}_{\tau}/\alpha_{\tau}$,
leaving behind a small difference that obeys the
Arrhenius law, $\exp[-E_{*}(g)\tau].$ }
\end{figure}

For $1\lesssim g < 4$, where perturbation theory in $g$ is not justified, we have computed $\mathcal{K}_{\tau}^{d}$ and $\mathcal{K}_{\tau}^{p}$
using a path integral Monte Carlo approach (see
Fig.\ref{f:numerics}). At sufficiently long times, $E_{c}^{*}(g)\tau\gg1,$
both these functions have anomalous $1/\tau^2$ dependences, but these cancel, leaving behind an exponentially small $\mathcal{K}_{\tau}.$

\begin{figure}
\includegraphics[%
  width=0.50\textwidth,
  keepaspectratio]{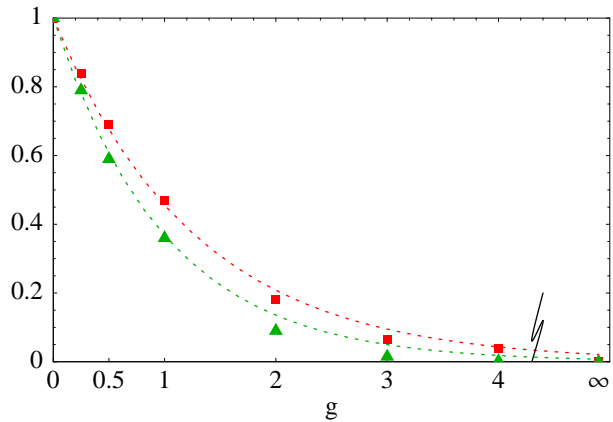}
\caption{\label{f:grankubo_estar} (Color online) Effective charging energies
  $E^{**}(g)$ (squares) and $E^{*}(g)$ (triangles), estimated from decay
  constants of $\mathcal{C}_\tau$ and $\mathcal{K}_\tau/\alpha_\tau$
  respectively.  Dotted lines are $e^{-\pi g/4}$ and $e^{-g}$.  At large $g$ ($\geq 3$), estimates suffer from error due to (i) discretization, (ii) insufficiently large $\beta$.}
\end{figure}

The numerical data show that $\mathcal{C}_\tau$ and $\mathcal{K}_\tau/\alpha_\tau$ decay exponentially, but the effective charging 
energies ($E^{**}$ and $E^{*}$ respectively) are reduced from $E_c$.  Fig.~\ref{f:grankubo_estar} shows our estimates of  $E^*$ and $E^{**}$.  The
site correlator has a gap that approximately obeys $E^{**}(g) \approx E_{c}e^{-\pi g/4}$. In our earlier papers\cite{tripathi04,loh05}, we have proposed that for a \emph{single} junction, the characteristic energy is $T_{*} \sim E_{c}e^{-\pi g/2}$. This has a simple physical interpretation. 
Delocalization of the charge over $N$ grains reduces the charging energy to
$E_{c}/N$, however the probability of correlating the phases on $N$
neighboring grains decreases exponentially as $p_{N} \sim p_1^N \sim
e^{-NT_{*}/T}$.  Optimizing with respect to $N$ gives the probability of
single-charge excitation for a granular chain, $p \sim e^{-2\sqrt{T_{*}E_{c}}}$, i.e., $E^{**}(g) = \sqrt{T_{*}E_{c}} \sim E_{c}e^{-\pi g/4}$. This result of ours for $E^{**}(g)$ would agree with the analysis in Ref.\cite{altland04} if the single-junction characteristic energy is taken to be $T_{*}\sim E_{c}e^{-\pi g/2}$ instead of $E_{c}e^{-\pi g}$ used in Ref.\cite{altland04}. The characteristic energy for a 1D granular chain thus differs from that inferred from the analysis in Ref.\cite{efetov02}. The effective Arrhenius gap for the conductivity, $E^*(g)$ fits well to $E^*(g)\approx E_{c}e^{-g}$ and differs from existing analytical predictions.  We cannot rule out numerical subtleties causing the difference between $E^{*}(g)$ and $E^{**}(g)$.

Renormalization group calculations (Refs.\cite{efetov03,falci95}) for a \emph{single electron box} show that the conductance $g > 1$ always renormalizes
to lower values as temperature is reduced. If the same statement holds for the AES model for the granular array, and there is no \emph{a priori} reason why it should, the implication is that the regular AES model has an Arrhenius conductivity at low temperatures regardless of how large $g$ is.

The physical picture emerging from our calculation is that \emph{for an ordered array}, sequential cotunneling processes which take charge around a loop
make anomalous contributions to certain intergrain correlation functions but \emph{not} to the conductivity.  The conclusion is that anomalous behavior in conductivity, such as soft activation or $T^{2}$, should not be automatically 
inferred from anomalous behavior in certain correlation functions or in the partition function.

We conclude with a discussion of disorder effects. 
Experiments on disordered 2D and 3D granular metals\cite{sheng73,chui81,simon87,gerber97} often reveal a `soft-activation' law $\sigma(T)\sim\exp[-\sqrt{T_{0}/T}]$.  More recently, the conductivity of carefully prepared nanoparticle arrays with controllable disorder \cite{beverly02} has been found to follow an Arrhenius law $\sigma(T)\sim\exp[-E^{*}/T]$ for {\it ordered} arrays, crossing over to soft-activation behavior with increased disorder in the position and size of the grains. Another relevant source of disorder is impurities in the insulating substrate which create random gate voltages $V_{g\mathbf{i}}$  at the grains.  
The presence of such disorder reduces the typical charging energy of the
grains; in particular, the charging energy vanishes when the grain background charge $Q_{0\mathbf{i}}= C_{\mathbf{i}}V_{g\mathbf{i}}= \frac{1}{2}$. This in turn makes energy levels available in the entire range $[-E_{c}/2,E_{c}/2]$. In combination with inelastic cotunneling, variable-range hopping can lead to a 
Mott or Efros-Shklovskii\cite{efros75} law for conductivity.  This differs from the original Efros-Shklovskii mechanism in that variable-range hopping due to wavefunction overlap is replaced by variable-length sequences of inelastic cotunneling events\cite{feigelman05}.   In our previous work\cite{tripathi04}, 
we recognized the importance of cotunneling over many grains in an eventual explanation for the soft-activation behavior.  However, as we have shown in this paper, cotunneling by itself is insufficient, and only Arrhenius conductivity is possible for arrays without disorder. 

It is also possible to create a range of energy levels in the interval
$[-E_{c}/2, E_{c}/2]$ through strong disorder in intergrain tunneling. In
particular, if in some part of the system, the intergrain tunneling is large,
then, as we have just shown, the charging energy is exponentially small,
$E^{**}(g)\sim E_{c}e^{-\pi g/4}$ for a chain and even smaller for higher
dimensional arrays. Since actual experiments are performed at not too low
temperatures $T > \delta \sim O(1K)$, we only need to make puddles available
up to charging energies as low as $E^{**}(g)=\delta$. 
This is plausible if tunneling disorder
is strong so that all conductances in the puddle are larger than
 $g > (4/z\pi)\ln(E_{c}/\delta)$. Once again, variable-range hopping arguments in combination with inelastic cotunneling will lead to a soft activation law. This could be an independent possibility in the position-disordered arrays in Ref.~\onlinecite{beverly02}.  In the presence of background charge disorder, even positionally regular arrays will show soft-activation behavior.   An explicit calculation of the conductivity for the disordered AES model will be taken up in a forthcoming work.  

YLL thanks Trinity College, Cambridge and the Cavendish Laboratory for
support. VT thanks Trinity College for a JRF.

\end{document}